# Approche Automatique de Génération des opérateurs ETL

Wided Bakari *
MIR@CL, Université de Sfax,
FSEGS, PB 1088, 3018 Sfax
Tunisia,
wided_bakari@yahoo.fr

Mouez Ali *
MIR@CL, Université de Sfax,
ISIMS, PB 242, 3021 Sfax
Tunisia,
mouez.ali@fmsf.rnu.tn

Hanene Ben-Abdallah *
MIR@CL, Université de Sfax,
FSEGS, PB 1088, 3018 Sfax
Tunisia,
hanene.benabdallah@fsegs.rnu.tn

*Résumé*

*Cet article aborde la génération des opérateurs ETL (Extract-Transform-Load) pour alimenter un magasin de données à partir d'une source de données relationnelle. Dans un premier temps, nous ajoutons de nouvelles règles à celles proposées par les auteurs de [1], ces règles traitent la combinaison d'opérateurs ETL. Dans un second temps, nous proposons une approche automatique à base de transformation de modèles pour la génération des opérations ETL nécessaires pour le chargement d'un entrepôt de données. Cette approche offre la possibilité au concepteur de définir certaines conditions nécessaires pour le chargement.*

*Mots clés : ETL, Extraction, Transformation, Chargement, sources de données, entrepôt de données, magasin de données, XSLT, opérateur.*

## I. INTRODUCTION

Dans les projets décisionnels, la phase de définition des opérateurs d'extraction, transformation et chargement (ETL-Extract-Transform-Load) des données dans l'entrepôt est souvent sous-estimée. Bien que ces tâches représentent (selon les spécialistes du domaine) à peu près les trois quart du projet décisionnel [13], les outils du marché n'offrent pas d'assistance. Le développeur est, par conséquent, contraint à comprendre aussi bien les sources d'alimentation que l'entrepôt en cours de construction. Une telle connaissance est nécessaire pour créer manuellement les correspondances entre la source de données et les éléments multidimensionnels.

L'ETL n'est pas un simple programme d'extraction, transformation et chargement de l'entrepôt de données et ne doit pas être traité de la sorte, mais il s'agit plutôt d'un système complexe qui permet d'offrir un environnement de développement, des outils de gestion des opérations et de maintenance. Le chargement de l'entrepôt de données à travers le processus ETL constitue une tâche difficile et qui prend beaucoup de temps pour sa réalisation avec un coût considérable dans les systèmes humains et les ressources financières.

La définition des opérateurs ETL a été étudiée, entre autre, dans les travaux de l'équipe de Systèmes d'Information Décisionnels du laboratoire Mi@cl, *c.f.* [1]. Ces travaux ont proposé une approche de génération des opérateurs ETL pour alimenter un magasin de données à partir d'une source relationnelle. Pour ce faire, ils prennent en entrée le schéma conceptuel du magasin de données, mis en correspondance avec une source relationnelle. Ils définissent un ensemble de règles pour générer les opérateurs ETL pour un SGBD relationnel. Ces travaux offrent la majorité des règles nécessaires mais d'autres règles ont dues être ajoutées.

L'objectif de ce travail est d'étudier les apports de la transformation de modèles pour le développement des entrepôts de données. En particulier, il vise à examiner l'applicabilité des techniques de transformation disponibles dans la phase de génération des procédures d'extraction-transformation-chargement des entrepôts de données.

Dans cet article, nous étendons les règles proposées par les auteurs de [1] par de nouvelles règles qui traitent d'autres combinaisons d'opérateurs. De plus nous mettons en œuvre cette proposition à travers une approche de transformation de modèles en utilisant le langage XSLT [9].

Le présent article est organisé en quatre sections, dont la première vise à présenter quelques approches de génération des opérateurs ETL. La deuxième section permet de présenter globalement notre démarche proposée pour la génération des opérateurs ETL. La troisième section permet de finaliser notre contribution par les nouvelles règles de génération des opérateurs ETL par rapport au travail de [1]. Finalement, cet article expose la conclusion et les perspectives de ce travail.

## II. ETAT DE L'ART

Dans la littérature de génération des opérateurs ETL, plusieurs approches ont été proposées. Parmi ces approches, on trouve plusieurs approches qui sont basées sur une ontologie dont l'utilisation ou la construction comme [2], [3], [4] et [5]. Par contre, il y a d'autres approches qui évitent la construction d'une telle ontologie, par exemple, l'approche proposée par les auteurs dans [1], pour la génération semi-automatique des procédures ETL en se basant sur une correspondance structurelle et sémantique entre la source de données et les éléments multidimensionnels.

Les auteurs de [2], présentent une approche basée sur les technologies du web sémantique pour faciliter le processus de sélection des informations pertinentes à partir des sources de données pour transformer ces données et les charger dans un entrepôt de données. Donc, les auteurs de [2] utilisent l'ontologie dans le but de spécifier la sémantique des schémas de la source de données ainsi que le schéma de l'entrepôt de données. Cette approche est adaptée pour capturer les informations nécessaires pour la conception des processus ETL. Elle est composée de quatre étapes qui sont : Définition du vocabulaire commun de l'application (ensembles de termes), Annotation des sources de données, Génération d'ontologie à partir du vocabulaire et de l'annotation des sources de données et Génération des opérateurs ETL. Par ailleurs, dans la génération des opérateurs ETL l'ontologie est réalisée manuellement.

Pour compléter le travail réalisé dans [2], les auteurs de [4] et [5] proposent une deuxième approche de construction de l'ontologie, cette approche est décomposée en trois étapes: d'abord, elle assure l'extraction manuelle de la sémantique des éléments de la source de données ainsi de l'entrepôt de données; puis, elle augmente les noms des concepts extraits par leurs synonymes et d'autres informations comme les différents formats représentés, les attributs agrégés, etc. Cette phase contient trois sous étapes : La construction de l'ontologie, annotation des sources d'alimentation et de l'entrepôt, et la génération des opérateurs ETL. La troisième étape est consacrée pour la génération des opérateurs ETL.

Par ailleurs, les auteurs de [3] proposent une approche de génération des opérateurs ETL en se basant sur l'utilisation d'une ontologie. Cette approche est composée de quatre principales phases dont la première détermine le processus d'extraction manuelle des données à partir des sources de données, ces données seront utilisées pour la construction de l'ontologie. A ce niveau, Il existe une ontologie globale et ontologie locale. La deuxième est la phase de mapping de l'ontologie dans laquelle définissent des relations sémantiques entre chaque ontologie locale avec celle globale. La troisième phase assure la dérivation de l'ontologie à partir des résultats de la phase précédente. La quatrième phase sert à définir manuellement les règles d'ETL afin d'opérer le processus ETL.

Bien que ces approches se basent sur une ontologie, elles ne s'intéressent pas à plusieurs types de relations sémantiques, comme la synonymie, hyperonyme, Meronyme, hyponyme et holonyme. De plus, nous constatons que ces approches raisonnent sur les ontologies pour dégager les opérateurs sans les générer. Par ailleurs, ces approches ne déterminent pas les enchaînements à suivre pour aboutir au chargement de l'entrepôt.

A ce niveau, les auteurs de [1] proposent une approche qui assure la génération semi-automatique des procédures ETL en se basant sur une correspondance structurelle et sémantique entre la source de données et les éléments multidimensionnels. Ainsi, les auteurs de [11] présentent un ensemble de règles qui permettent la génération des procédures ETL. Pour atteindre cet objectif, les auteurs de [11] prennent en compte le mapping qui relie chaque terme du magasin de données à un ou plusieurs termes de la source de données en précisant la relation sémantique entre eux (Synonyme, Hyperonyme, Hyponyme, Meronyme,…).

### III. APPROCHE PROPOSEE POUR LA GENERATION DES OPERATEURS ETL

Pour la génération des opérateurs, nous proposons une approche qui se base sur la transformation de modèles. Donc, la génération des opérateurs ETL se base essentiellement sur la correspondance entre la source et la cible de données. Dans notre cas, nous parlons de la correspondance entre les éléments de l'entrepôt et ceux de la source de données. Cette correspondance est apparue dans le fichier de départ qui est le Mapping [1]. Notre approche détermine des scénarii de chargement des magasins de données. Pour assurer la génération des opérateurs ETL, on doit tenir compte des tâches suivantes : La génération des opérateurs ETL, les règles de transformation, le processus de transformation de modèles et de chargement du modèle cible.

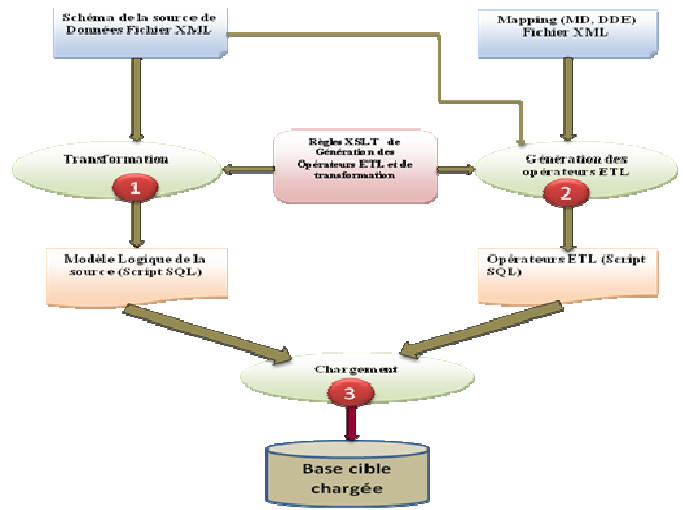

Figure 1: Démarche de génération des opérateurs ETL

Comme illustré dans la Figure 1, notre démarche de génération des opérateurs ETL prend en entrée deux fichiers XML : le premier c'est le Mapping [1] qui décrit pour chaque élément du magasin de données ses correspondances avec les éléments de la source. Le schéma multidimensionnel du magasin de données présenté dans la figure 2 indique le fait, ses mesures et ses dimensions. De plus, pour chaque dimension, il indique l'ensemble des paramètres ainsi que leurs attributs faibles. Le deuxième fichier d'entrée contient le schéma de la source de données. Ceci décrit des informations concernant les tables, leurs attributs, les contraintes de clé primaire et étrangère, etc. Ce schéma est utilisé pour définir le modèle logique de la source de données. Ce dernier est utilisé pour la génération des opérateurs ETL.

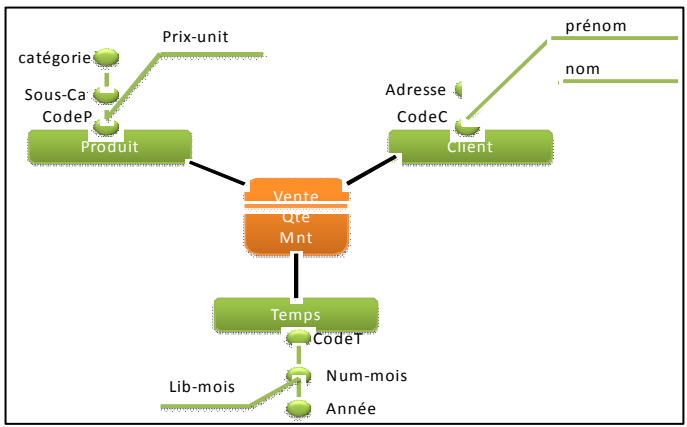

Figure 2: Schéma multidimensionnel du magasin de données Vente

Le mapping est obtenu (semi-automatiquement) à partir du schéma multidimensionnel ainsi que le DDE de la source d'alimentation. Il décrit pour chaque élément du magasin de données les relations sémantiques qu'il a avec les éléments de la source d'alimentation. Il est dérivable grâce à l'application des règles de conception de magasin de données par une approche ascendante [12].

Par ailleurs, le DDE [8] s'agit d'un un dictionnaire qui contient toutes les informations du méta-schéma de la source de données (les tables, leurs attributs, les contraintes de clé primaire et étrangère,…). Ces informations peuvent être automatiquement extraites à partir du SGBD de la source. De plus, ces informations sont étendues par des descriptions textuelles et une liste de synonymes, hyperonymes et hyponymes pour chaque élément de la source.

Les règles de génération des opérateurs ETL sont dégagées en utilisant le Mapping [1] et en s'appuyant sur les relations sémantiques définies entre chaque élément du magasin de données avec chaque élément de la source. Les règles pour réaliser cette transformation ont été développées avec le Language XSLT [9]. Ce dernier est un langage de transformation qui prend en entrée des documents XML.

La génération des opérateurs ETL nécessite : la compréhension des sources de production et de l'entrepôt à développer, la définition des correspondances structurelles et sémantiques entre les deux types de bases de données [1], pour enfin la définition des opérateurs d'extraction, de transformation et de chargement nécessaires pour alimenter l'entrepôt de données.

Les résultats d'application de nos règles XSLT, sont les opérateurs ETL utilisés pour charger un modèle cible à partir du modèle source. Dans notre cas on parle des scripts SQL (l'ordre de SQL SELECT et SQL INSERT). Ces derniers permettent le chargement des dimensions et du fait à partir des tables sources correspondantes. Nous avons validé notre approche par une étude de cas « vente de produits ».

Dans la section 4, nous illustrons quelques exemples de règles que nous avons ajoutées, ainsi, les cas d'application de ces règles.

IV. REGLES DE GENERATION DES OPERATEURS ETL

Pour la génération des opérateurs ETL, nous utilisons les opérateurs de base résumés dans la TABLE I. De plus, nous adoptons les notations suivantes:

- *Mapping* : le fichier Mapping contenant les correspondances.
- *ETL* : L'opérateur ETL à appliquer.
- *Card* : la cardinalité (nombre d'attributs).
- *Dim* : Une dimension.
- *TS* : Une table source.
- *F* : Un fait.
- *Parametre* : Un paramétre.
- *AttF* : Un attribut faible.
- *Att* : Un attribut source.
- *Mesure* : Une mesure.
- *C* : une condition fixée par le décideur (condition de filtrage, fonction d'agrégation, format de conversion…). Notons que C prend plusieurs formats comme OP (valeur).

Avec OP= {>, <,>=, <=,=, <>, like} et valeur= (numérique ou chaine de caractère). C'est une combinaison de plusieurs conditions (du format précédent) séparées par les opérateurs logiques « && » et « || ».

- *R= {synonymie, Hyperonyme, Hyponyme, Holonyme, Meronyme}* : un ensemble de relations entre les paramétres ou attributs faibles ou mesures avec les attributs sources.

Pour appliquer nos règles, considérons le schéma logique de la source suivante:

**Produit (codeP, description, prix-unit, codecat#)**

**Sous_Catégorie (codesouscat, désignation, codecat#)**

**Catégorie (codecat, designation)**

**Client (codeC, nom, prénom, ville, codepostale)**

**Facture (refF, dateF, codeC#)**

**Lignes_fact (refF#, codeP#, quantite, montant)**

TABLE I. OPÉRATEURS ETL UTILISÉS [1]

| Opérateur | Description |
|---|---|
| $\pi(a1, a2,...RS)$ | La projection conserve les attributs intéressants (a1, a2,...) de RS pour le chargement du magasin de données |
| $C(a1, a2,..., b)$ | Concaténation des attributs a1, a2 etc. pour avoir la valeur de l'attribut b |
| $S(a, b1, b2,...)$ | Diviser la valeur de l'attribut a sur les attributs b1, b2. |
| $FC(a, F1, F2)$ | Convertir la valeur de l'attribut a du format F1 vers le format F2 |
| $\gamma(a1, a2,..., fonction\ d'aggrégation)$ | Les fonctions d'agrégation permettent de faire des opérations sur une sélection de champ |
| $\delta(a1, a2,..., condition)$ | Conserver les tuples qui respectent une condition définie sur les valeurs des attributs a1, a2 etc. |
| $Nn\ (att)$ | Eliminer les valeurs nulles |
| $JOIN$ | Elle consiste à juxtaposer les tuples dont la valeur d'un attribut est égale dans les deux tables. |

**Règle 1 : Pour** toute *Dim* du *Mapping* **faire**

    Pour toute *R* faire
    Pour tout *Parametre* (ou *AttF*) de Dim Tel que Mapping(*Dim.Parametre*)={(*TS*.Att1, Hyponyme), (*TS*. Att2, Hyponyme) …} faire
        ETL(*Dim.Parametre*)=**C**(**δ**(TS.Att1:C1, *TS*.Att2:C2), *Parametre*).
    **Fin pour**

Le chargement du paramètre « adresse » de la dimension « client » correspond à une combinaison de deux attributs sources « codepostale » et « ville » de la table source « client » à l'existence d'une relation d'hyperonymie entre la source et la destination. Donc, d'après la règle 1 pour charger ce paramètre, il faut tout d'abord, appliquer des conditions de filtrage définies par le décideur sur les deux attributs sources « codepostale » et « ville ». Puis, d'assurer une concaténation sur le résultat de cette sélection.

Dans le cas inverse, où la dimension « client » contient les paramètres suivants « codepostale » et « ville » et la table source « client » admet un attribut source « adresse ». Donc, à ce niveau il faut diviser la valeur de l'attribut source « adresse » selon une condition précisée par le décideur en deux valeurs, une pour le « codepostale » et autre pour la « ville ». Puis charger ces valeurs respectivement dans la dimension en tant que « codepostale » et « ville ». Ce cas sera traité par la règle 2.

**Règle 2 : Pour** toute *Dim* du *Mapping* **faire**
    Pour tout *Parametre* (ou *AttF*) de *Dim* Tel que Mapping (*Dim.Parametre*) = {(*TS*.Att; Hyponyme)} faire
    ETL (*Dim.Parametre*) = **S** (**δ** (*TS*.Att, C*), att1, att2, att3...).
    **Fin pour**

Lorsqu'un paramètre (Attribut faible) d'une dimension ou une mesure d'un fait possède les mêmes valeurs que celles d'un attribut source sauf que le décideur définit une valeur particulière pour ce paramètre (attribut faible) ou mesure. Dans ce cas, on applique l'opérateur select (**δ**) pour convertir les valeurs de l'attribut source selon une condition précisée par le décideur. La Règle 3 traite ce type de cas.

**Règle 3 : Pour** toute *Dim* du *Mapping* **faire**
    Pour tout *Parametre* (ou *AttF*) de *Dim* Tel que *Mapping* (*Dim.Parametre*)={(*TS*.Att; Synonymie)} faire
      **Si** (*C* = Format_Convert (*TS*.att, F1, F2))
    **Alors**
      ETL (*Dim.Parametre*) = **FC** (**δ** (*TS*.Att, C*), F1, F2).
    **Fin pour**

Un autre cas possible, si un décideur peut appliquer une fonction d'agrégation sur l'attribut source « quantite » de la table source « lignes_fact », dans le but de charger cette valeur dans la mesure « qte » du fait « vente », comme il peut modifier le format du résultat obtenu (ex. numérique → chaine). Ce cas est traité par la règle 4 ou il ya combinaison de trois opérateurs Format_Convert, fonction agrégation et Project.

**Règle 4 : Pour** toute *Dim* du *Mapping* **faire**
    Pour tout *Parametre* (ou *AttF*) de *Dim* Tel que *Mapping* (*Dim.Parametre*)={(*TS*.Att; Synonymie)} faire
      **Si** (C1) et (C2 (*TS*.att, F1, F2)) et (C3)
    **Alors**
      ETL (*Parametre*) = **γ** (**FC** (**π** (*TS*. Att ...), F1, F2), fonction d'agrégation)
    **Fin pour**

Dans l'exemple traité, si on a une table source « vente-produit » qui contient un attribut source appelé « prix_total » et qu'on a un fait « prix_prod » ayant une mesure « prix_unit », avec l'attribut source « prix_total » est le résultat d'une formule mathématique : prix-total=quantite*prix-unitaire. Alors, prix_unit = prix_tolal / quantite. Le chargement de la mesure « prix_unit » peut être assuré en appliquant les trois opérateurs fonction d'agrégation, Project et Join. Ceci sera traité par la règle 5.

**Règle 5 : Pour** toute *Dim* du *Mapping* **faire**
    Pour tout *Parametre* (ou *AttF*) de *Dim* Tel que *Mapping* (*Dim.Parametre*)={(*TS*.Att; Synonymie)} faire
      **Si** (C1) et (C2 (*TS*.att, F1, F2)) et (C3)
    **Alors**

ETL (*Parametre*) = $\gamma$ ($\delta$ (**Join** (TSi, TS1), TS.Att, C), fonction d'agrégation).
**Fin pour**

## V. PROTOTYPE

Afin d'évaluer les apports et les limites de notre démarche de génération des opérateurs ETL, nous avons développé un environnement pour assister le décideur du magasin de données lors de la définition aussi bien des opérateurs ETL. Cet environnement, développé avec Java et appliquant les règles qui sont déjà proposées par les auteurs de [1], et nos règles ajoutées qui sont déjà présentées dans la section 4. De même, cet environnement assure la génération des opérateurs ETL nécessaires pour le chargement des dimensions et des faits. L'environnement développé nous a permis de mener divers évaluations expérimentales.

Le chargement de magasin de données nécessite toujours d'indiquer la source au prés de laquelle le système extrait les informations, les transformes, par la suite, les stocke dans le magasin correspondant. Le prototype permet au concepteur ETL de visualiser le schéma de la source et celui du magasin. Ainsi, il visualise les données de la base source. La figure 3 présente la source de données.

Figure 3: Sources de données

Ce prototype affiche, pour chaque hiérarchie d'une dimension, les opérateurs ETL pour l'ensemble de ses paramètres et ses attributs faibles. Ainsi, il affiche les opérateurs ETL pour les mesures d'un fait.

Figure 4: Les opérateurs ETL pour les paramètres et les attributs faibles de l'hiérarchie H2 de la dimension Produit

Après avoir affiché la source de données et le magasin à charger, le système assure la phase de transformation pour générer les opérateurs ETL pour chaque paramètre et chaque attribut faible de toutes les dimensions d'un fait. Cette étape présente le processus de transformation de modèles en appliquant nos règles XSLT [9] sur le fichier d'entrée XML.

Figure 5: Script SQL généré pour les opérateurs ETL de la Figure 4.

Une fois, la source et la destination ont été choisies, le système génère les opérateurs ETL pour le fait et ses dimensions. Ce qui reste à faire est l'insertion ou le chargement des données source dans le magasin de données.

Figure 6: Résultat du chargement du fait et de ses dimensions : version relationnelle.

La figure 6 illustre un cas de chargement relatif à la dimension « client ». Ainsi, l'opération de chargement est réalisée en appliquant une simple projection (opérateur *Project*) sur les attributs sources « codec », « nom » et « prenom » de la table source « client », et en appliquant une concaténation (opérateur *Concat*) sur les attributs sources « ville » et « codepostale » pour que le système récupère les informations du paramètre « adresse ». La figure 7 montre la version multidimensionnelle du résultat de ce chargement.

| | DIMENSION_KEY | NOM | NIVEAU_CODE... | NIVEAU_CODE... | PRENOM | NIVEAU_ADRESS... | ADRESSE | NIVEAU_ADRES... |
|---|---|---|---|---|---|---|---|---|
| 1 | -11 | MoufRUE | -11 | 13 | KHALID | | | |
| 2 | -12 | BBB | -12 | 5 | BBB | | | |
| 3 | -13 | zAkl | -13 | 12 | ANASS | | | |
| 4 | -14 | AAA | -14 | 3 | BBB | | | |
| 5 | -15 | Ali | -15 | 1 | mohamed | | | |
| 6 | -16 | ELHASSAK | -16 | 10 | MUSTAPHA | | | |
| 7 | -17 | WAJDI | -17 | 2 | IMEN | | | |
| 8 | -18 | Esbaiss | -18 | 11 | MOUAD | | | |
| 9 | -19 | MOHAMED | -19 | 4 | NOUR | | | |
| 10 | -20 | hadri | -20 | 14 | ZAKARIA | | | |
| 11 | 31 | | | | | 31 | SFAX3000 | 13 |
| 12 | 32 | | | | | 32 | SOUSSE5000 | 5 |
| 13 | 33 | | | | | 33 | TUNIS1000 | 12 |
| 14 | 34 | | | | | 34 | TUNIS1000 | 3 |
| 15 | 35 | | | | | 35 | sfax3000 | 1 |
| 16 | 36 | | | | | 36 | GAFSA2100 | 10 |
| 17 | 37 | | | | | 37 | GAFSA2100 | 2 |
| 18 | 38 | | | | | 38 | SFAX3000 | 11 |
| 19 | 39 | | | | | 39 | sfax3000 | 4 |
| 20 | 40 | | | | | 40 | GAFSA2100 | 14 |

Figure 7: Résultat du chargement de la dimension client à partir de la table source client : version multidimensionnelle.

VI. CONCLUSION ET FUTURS TRAVAUX

Dans cet article, nous avons proposé de nouvelles règles de transformation XSLT pour générer des opérateurs ETL dans le but de charger un entrepôt de données. Ce travail est motivé par le fait que l'une des tâches importantes à réaliser durant le développement d'un entrepôt de données est la génération du code à partir du modèle source pour assurer le chargement de l'entrepôt de données à partir des sources d'alimentation. Afin d'étudier tous les cas possibles dans le chargement de l'entrepôt de données en vue de gérer des opérateurs ETL, nous avons proposé une approche basée sur la transformation de modèles pour la génération automatique des opérateurs ETL. En outre, nous avons présenté un prototype facilitant l'application de notre approche. Ceci nous permettra de mener des évaluations expérimentales, d'une part, pour valider nos règles et, d'autre part, pour déterminer une approche d'optimisation de leur application pour un meilleur passage à l'échelle.

Bien que nous estimions que les objectifs visés dans ce travail soient relativement atteints, Comme perspectives de ce travail, nous citons:

- Nous compterons réaliser une implémentation de notre approche dans d'autres outils de transformation pour comparer les performances et essayer de tirer profit des avantages qu'ils offrent.
- Nous compterons également réaliser des transformations au niveau méta-modèles et faire une étude approfondie dans ce domaine.
- Etendre formellement cette approche pour qu'elle soit générale et ne se limite pas à un fichier d'entrée spécifique.
- Utiliser plusieurs fichiers sources pour élargir notre système de génération des opérateurs ETL.